\documentclass[british,english,a4paper,reprint,superscriptaddress,aps,twocolumn]{revtex4}

\usepackage{graphicx,color}
\usepackage{amssymb,amsmath}

\newcommand{\avd}{\omega_\text{drive}}
\newcommand{\tod}{\tau_\text{drive}}
\newcommand{\vhi}{\varphi}
\newcommand{\x}{\mathbf r}
\newcommand{\dd}{\text{d}}
\newcommand{\dbar}{\mbox{\fontencoding{T1}\selectfont\dj}}
\newcommand{\mean}[1]{\langle #1\rangle}
\newcommand{\pecl}{\operatorname{\mathit{P\kern-.08em e}}}

\begin{document}

\title{Transmission of torque at the nanoscale}

\author{Ian Williams\footnote{Present Address: Department of Chemical Engineering, University of California Santa Barbara, Santa Barbara, California, 93106, USA}}
\affiliation{H.H. Wills Physics Laboratory, Tyndall Ave., Bristol, BS8 1TL, UK}
\affiliation{School of Chemistry, Cantock's Close, University of Bristol, BS8 1TS, UK}
\affiliation{Centre for Nanoscience  and Quantum Information,
Tyndall Avenue, Bristol BS8 1FD, UK}
\author{Erdal C. O\u{g}uz}
\affiliation{Department of Chemistry, Princeton University, Princeton, New Jersey, 08544, USA}
\affiliation{Institut f\"ur Theoretische Physik II, Heinrich-Heine-Universit\"at, D-40225 D\"usseldorf, Germany}
\author{Thomas Speck}
\affiliation{Institut f\"{u}r Physik, Johannes Gutenberg-Universit\"{a}t Mainz, Staudingerweg 7-9, D-55128 Mainz, Germany}
\author{Paul Bartlett}
\affiliation{School of Chemistry, Cantock's Close, University of Bristol, BS8 1TS, UK}
\author{Hartmut L\"owen\footnote{e-mail: hlowen@hhu.de}}
\affiliation{Institut f\"ur Theoretische Physik II, Heinrich-Heine-Universit\"at, D-40225 D\"usseldorf, Germany}
\author{C. Patrick Royall\footnote{e-mail: paddy.royall@bristol.ac.uk}}
\affiliation{H.H. Wills Physics Laboratory, Tyndall Ave., Bristol, BS8 1TL, UK}
\affiliation{School of Chemistry, Cantock's Close, University of Bristol, BS8 1TS, UK}
\affiliation{Centre for Nanoscience and Quantum Information,
Tyndall Avenue, Bristol BS8 1FD, UK}

\date{\today}

\begin{abstract}
\textbf{
In macroscopic mechanical devices torque is transmitted through gearwheels and clutches. In the construction of devices at the nanoscale, torque and its transmission through soft materials will be a key component. However, this regime is dominated by thermal fluctuations leading to dissipation. Here we demonstrate the principle of torque transmission for a disc-like colloidal assembly exhibiting clutch-like behaviour, driven by $27$ particles in optical traps. These are translated on a circular path to form a rotating boundary that transmits torque to additional particles confined to the interior. We investigate this transmission and find that it is determined by solid-like or fluid-like behaviour of the device and a stick-slip mechanism reminiscent of macroscopic gearwheels slipping. The transmission behaviour is predominantly governed by the rotation rate of the boundary and the density of the confined system. We determine the efficiency of our device and thus optimise conditions to maximise power output. 
}
\end{abstract}

\maketitle

Classical thermodynamics evolved in response to the need to understand, predict, and optimise the steam engines responsible for driving the industrial revolution~\cite{carnot1824}. In contrast to these macroscopic devices, ``soft'' engines at the nanoscale operate in the presence of thermal fluctuations. When the thermal energy is of the same order as the work done, these fluctuations pose a fundamental challenge and call for new design principles. Nanomachines have been studied theoretically~\cite{seifert2011}, particularly in the case of molecular motors~\cite{parmeggiani1999,kolomeisky2013}.  Experimentally, colloidal and nanoparticle systems provide insight into fundamental thermodynamic processes in the presence of stochastic Brownian noise \cite{babic2005,pesce2011}. Single colloidal particles manipulated by optical forces have been used to \emph{e.g.} realise the analog of a Stirling engine~\cite{blickle2012}, to explore the nanoscopic manifestation of the second law of thermodynamics~\cite{carberry2004,gieseler2014} and to emulate memory devices testing Landauer's principle for the work dissipated when erasing information~\cite{berut2012}. The next stage is to exploit these insights from model systems to engineer devices that perform predictably in the presence of thermal fluctuations.

\begin{figure*}[htb]
	\begin{center}
		\centerline{\includegraphics[width=120mm]{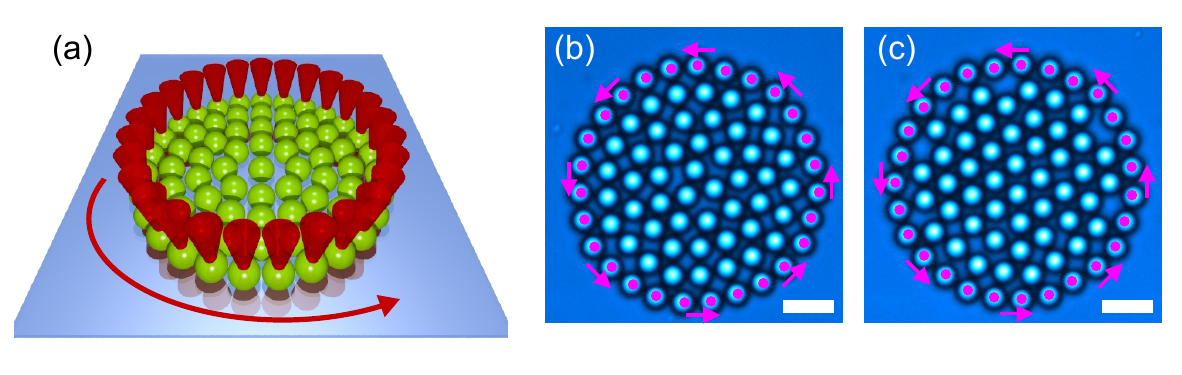}}
		\caption{\label{figSchematic} \textbf{The experimental system.} (a) Schematic of a model nano-clutch. Red cones represent optical traps translating $27$ particles on a circular path. The interior is populated with $N=48$ particles and assumes a bistable state, either (b) layered fluid, $\psi_6 \lesssim 0.8$, or (c) locally hexagonal, $\psi_6 \gtrsim 0.8$. Particles marked with magenta dots are optically trapped and translated in the direction indicated by the arrows. Scale bars represent $10 \; \mathrm{\mu m}$.}
	\end{center}
\end{figure*}

An important step in this direction would be a microscopic gearbox or transmission system. While rotational devices have been fabricated~\cite{klapper2010,asavei2013,lin2012}, and nanoscopic gearwheels realised~\cite{galajda2001,dileonardo2010}, such devices are typically single rigid objects driven to rotate by \textit{e.g.} optical forces~\cite{asavei2013,lin2012,galajda2001}, magnetic forces~\cite{xia2010}, rectified bacterial motion~\cite{dileonardo2010} or asymmetric catalytic activity~\cite{fournierbidoz2005,gibbs2009}. However, nanoscale devices are frequently engineered using soft components, and the self- or directed assembly of nanometric building blocks into soft mesostructures represents an exciting opportunity for the realisation of microscopic machines~\cite{whitesides2002,kim2014,puigmartiluis2014,ismagilov2002,grzelczak2010,kraft2012,sacanna2010}. The response of a soft material to an external force is fundamentally different from that of a rigid body and the transmission of torque through soft materials remains little explored despite its clear importance if nanoscale mechanical devices are to be developed. Since nanoparticle, colloidal and biological systems are typically suspended in a fluid medium, it is natural to investigate the effect of such a medium on a soft rotational device. It is clear from considerations of the effects of inertia in small bodies that the influence of the immersing fluid will be profound~\cite{purcell1977}.

Here we tackle the transmission of torque in soft devices. We realise a driven colloidal assembly exhibiting clutch-like behaviour and identify new mechanisms important to the transmission of torque through soft materials at the nanoscale. Our device is created in a quasi 2d colloidal system in which $n=27$ particles are held in a circular configuration of radius $R$ using holographic optical tweezers~\cite{bowman2013}. The interior region is populated with $N$ identical (but untweezed) particles creating a circularly confined system, as shown in Fig. 1 ~\cite{williams2014}.  The boundary is rotated to provide torque to the assembly. Details are provided in the Methods. The speed of this rotation is characterised by the P\'{e}clet number, $\pecl$, defined in the Methods, with a larger $\pecl$ indicating faster rotation. 

\begin{figure*}[htb]
	\begin{center}
		\centerline{\includegraphics[width=170mm]{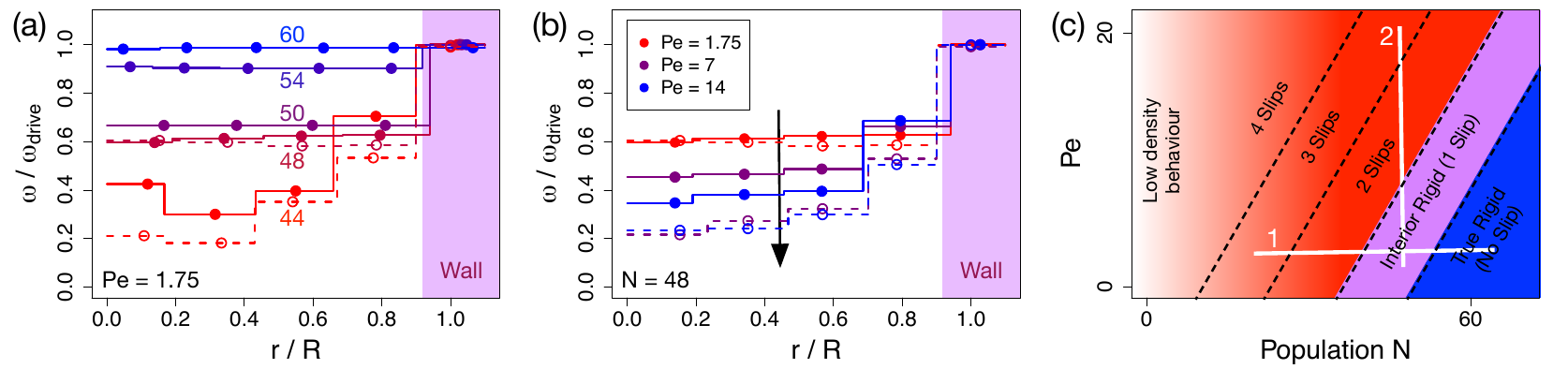}}
		\caption{\label{figPhasediagram} \textbf{Controlling rotational behaviour with $N$ and $\pecl$.} (a) Normalised angular velocity profiles for varying $N$ measured in simulation (solid data) and experiment (open data) at $\pecl=1.75$. Labels indicate $N$. Purple region is the boundary. (b) Normalised angular velocity profiles for varying $\pecl$ measured in simulation (solid points and lines) and experiment (open points and dashed lines) at $N=48$. Arrow indicates direction of increasing $\pecl$. Purple region is the boundary. (c) Schematic nonequilibrium state diagram showing behaviour in terms of the number of slips as a function of $N$ and $\pecl$. White lines indicate position of data series shown in (a), $1$, and (b), $2$.}
	\end{center}
\end{figure*}

In order to understand the key aspects of the dynamical behaviour, we perform Brownian dynamics simulations with parameters matched to the experiments.  In the experiments, the driven boundary exerts forces on the solvent, creating a fluid velocity field affecting the motion of the confined population. In simulation we determine an effective solvent flow field by superposing the flow fields due to each of the $n=27$ driven particles. Full simulation details are provided in the Methods.

Altering the radius of the driven boundary \textit{in situ} allows torque transmission to be engaged or disengaged at will forming a minimal model of a ``nanoclutch''. We investigate different regimes of torque transmission to the centre of the assembly dependent upon P\'{e}clet number and the density of the confined population. Due to the softness of the colloidal system, under certain conditions the rotational behaviour is markedly different from that of a rigid body. We consider the consequences of this for the efficiency of such systems. Our analysis reveals periodic structural self-similarity of geometric origin, related to slip between particle layers, leading to clutch-like behaviour due to a decoupling between driven and loaded parts of the device.

\section{Rotational behaviour}

Soft materials, such as assemblies of colloids, are much more susceptible to external perturbation (in the case of shear) than macroscopic solid bodies. This opens the possibility to exploit the mechanical properties of the material to modify the transmission itself. We first consider the system in its quiescent state. At sufficient density ($N \ge 47$), the static system exhibits a bistability between locally hexagonal and layered fluid configurations as shown in Fig. 1 (b) and (c)~\cite{williams2013,williams2014}. Hexagonality is quantified using the average bond-orientational order parameter $\psi_6$ (defined in the Methods). Without boundary rotation, both hexagonal and layered configurations are solid on the experimental timescale --- the structural relaxation time exceeds the experimental duration.

Upon rotation, we observe multiple modes of transmission dependent on both the confined population and $\pecl$, identified by qualitatively distinct angular velocity profiles. Figure 2 (a) shows the effect of varying $N$ at $\pecl=1.75$, where the angular velocity is averaged within particle layers and plotted with a point at the layer centre of mass. For $N=44$ (red data) the angular velocity profile has a step-like form with sharp discontinuous changes indicating slipping between circular layers rotating with different angular velocities. Here, in both simulation (solid red line) and experiment (dashed red line) one can clearly identify three slip locations between layers. On increasing the interior population to $N=48$ (red-purple lines) only a single slip is observed, occurring at $r/R \approx 0.9$, between the driven boundary and the confined population. Inside the wall the angular velocity profile is flat, indicating rigid-body-like rotation of the interior. Such interior rigid behaviour persists in simulations of populations up to $N=54$, albeit with the degree of slip decreasing on increasing $N$. By $N=60$ (blue line) there is no longer any slip between the boundary and the interior resulting in a flat angular velocity profile at $\omega(r) = \omega_{\mathrm{drive}}$ characteristic of full rigid body rotation. Note that at the greatest populations studied in simulation ($N=54$ and $N=60$) a structural rearrangement must occur resulting in the formation of a fifth particle layer and an outwards displacement of the wall particles.

Holding the population constant at $N=48$ and increasing $\pecl$ also results in the development of slip ring. This is illustrated in Fig. 2 (b). At $\pecl=1.75$ both experiment (open red data) and simulation (solid red data) exhibit interior rigid rotation, characterised by a flat angular velocity profile at $\omega \approx 0.6 \; \omega_{\mathrm{drive}}$ in the interior of the device. By $\pecl = 7$ a second slip has developed in both experiment and simulation.

Figure 2 (c) shows the non-equilibrium state diagram in terms of $N$ and $\pecl$. Rotational behaviour is characterised by the number of slip rings. Taking an initially rigid system and either increasing the driving speed or decreasing the population results in the sequential development of slip rings. This starts with the boundary slipping over the interior (interior rigid) and propagates inwards.  Slips develop between adjacent particle layers and thus their radial location is determined by the structure of the confined assembly. When altering $N$, small changes in slip locations are observed as the layered structure shifts to larger $r$ at higher density. At constant population, the radial slip locations remain unchanged as $\pecl$ is increased.

The line labelled ``1'' in Fig. 2 (c) represents the approximate location of the constant $\pecl$ data series presented in (a) while ``2'' represents the constant $N$ series shown in (b). Video examples of multiple slip and interior rigid behaviours from experiments with $N=48$ are available as Supplementary Movies 1 (multiple slip) and 2 (interior rigid). Simulated data showing full rigid body rotation can be seen in Supplementary Movie 3. 

\begin{figure}[htb]
	\begin{center}
		\centerline{\includegraphics[width=80mm]{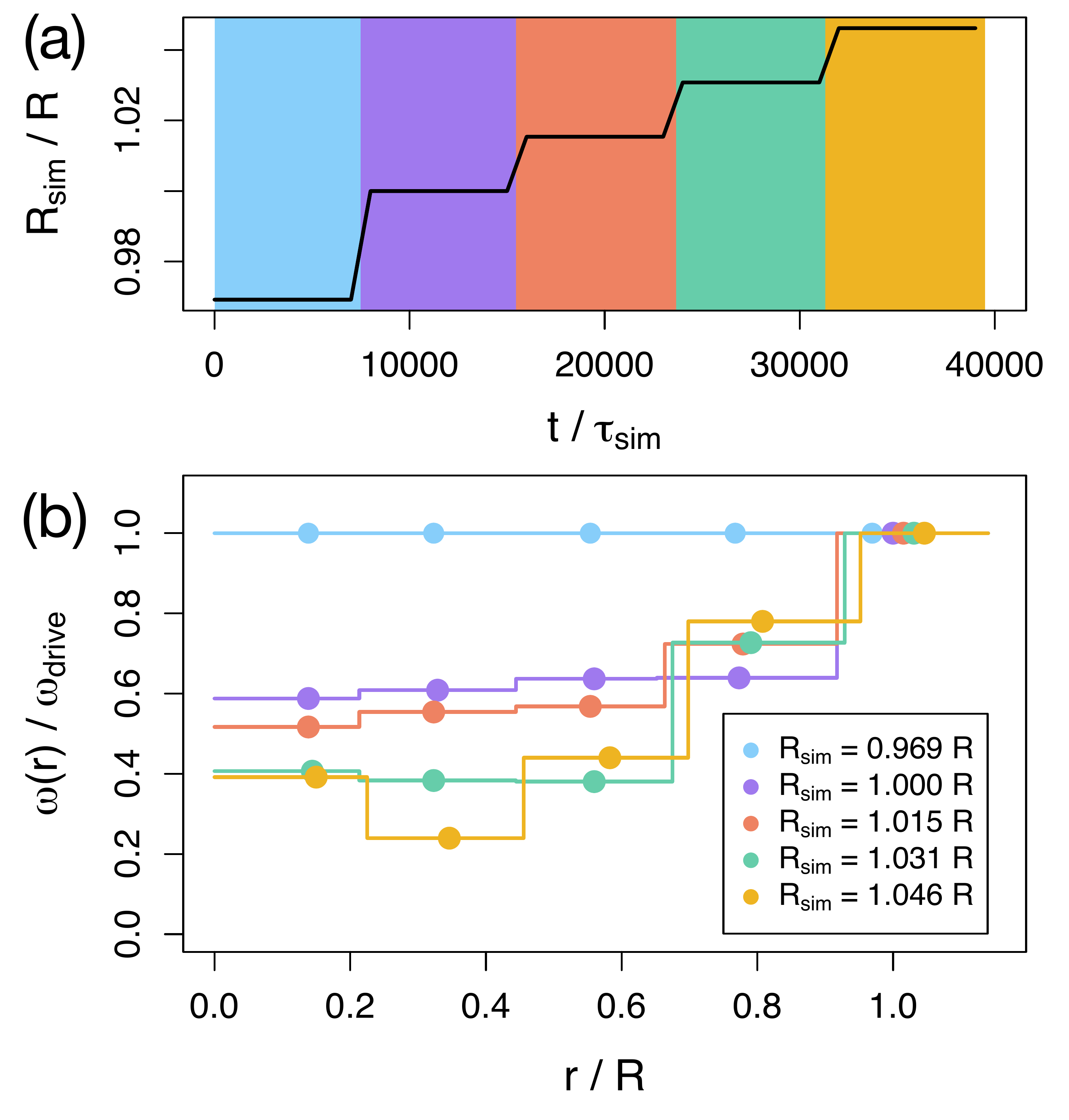}}
		\caption{\label{figChangingR} \textbf{Disengaging transmission \textit{in situ} by altering density in a single simulation with $N=48$.} (a) Radial location of boundary as a function of time in simulation time units. $R$ is the radius of the experimental system while $R_{\mathrm{sim}}$ is the simulated radius, given in terms of $R$. (b) Angular velocity profiles in the five simulated time intervals indicated by the colour shading in (a). }
	\end{center}
\end{figure}

Exploiting the dependence of rotational behaviour on $N$ allows clutch-like operation by altering the internal density \textit{in situ}. Figure 3 illustrates this in a single simulation at $N=48$ in which the radial location of the boundary is increased during the simulation, reducing the internal density. On increasing this radius from $R_{\mathrm{sim}} = 0.969 R$ to $R_{\mathrm{sim}} = 1.046 R$, slips develop between layers. Thus we demonstrate that not only does the rotational behaviour depend upon the internal density but that this effect can be employed as an active control mechanism, engaging or disengaging a microscopic rotational device at will in a clutch-like operation mode. In Supplementary Fig. S1 and Supplementary Note I, we show that \textit{in situ} alteration of $\pecl$ allows similar control of rotational behaviour. Thus, clutch-mode operation can be realised through variations in both driving speed and internal density, or combinations thereof, effectively ``dialling in'' the desired behaviour.

While we focus on the system with a boundary of $n=27$ particles resulting in a confined assembly of $4$ particle layers for all but the largest populations, simulations are also performed for smaller, $n=21$, and larger, $n=33$ boundaries with $3$ and $5$ layers respectively. Angular velocity profiles measured in these systems, analogous to those in Fig. 2 (a) and (b), are shown in Supplementary Fig. S2 and described in Supplementary Note II. In both these systems the same phenomena are observed with the same dependence on $N$ and $\pecl$, indicating that the mechanisms revealed here are not specific to the $4$-layer device geometry.

\begin{figure*}[htb]
	\begin{center}
		\centerline{\includegraphics[width=160mm]{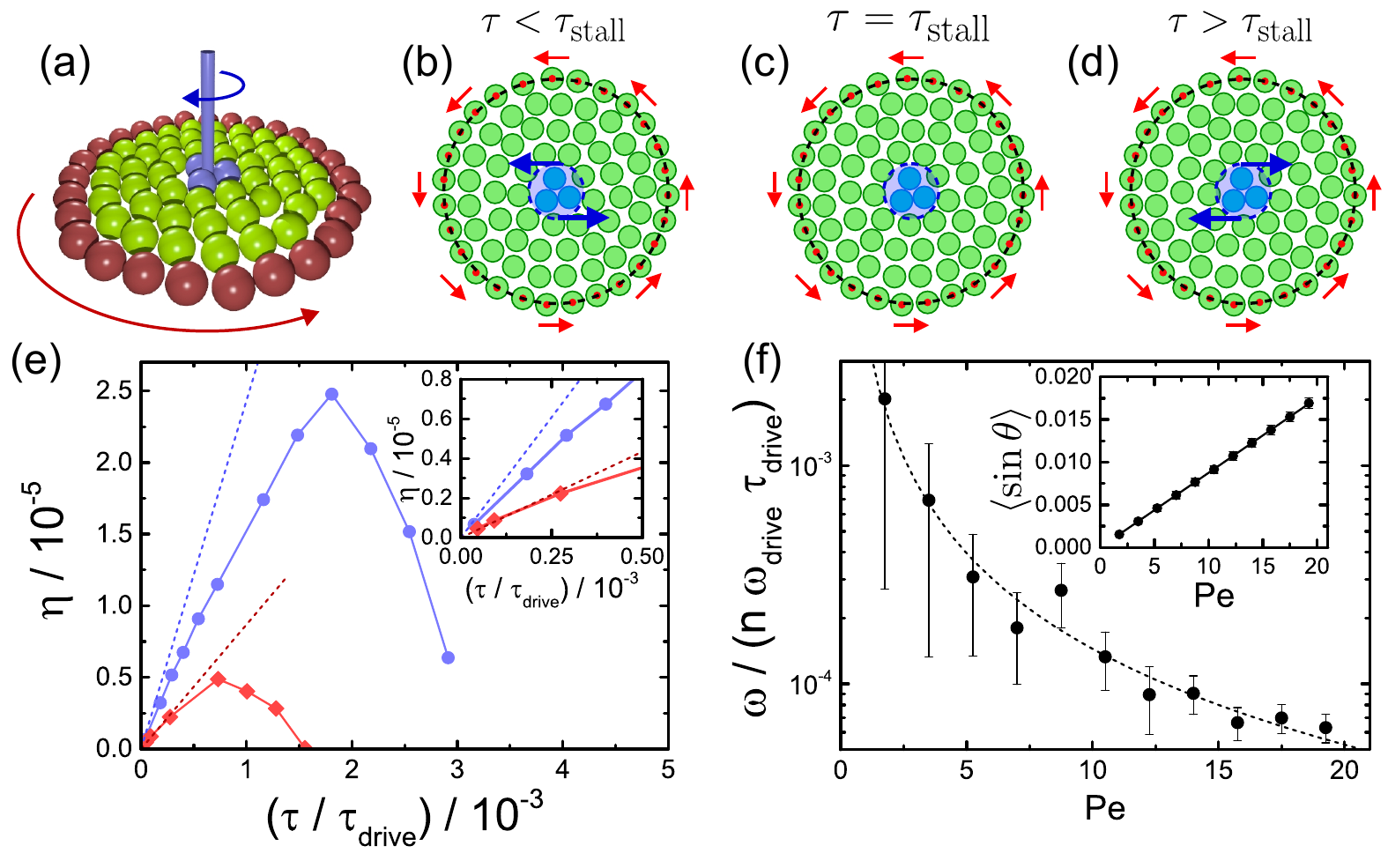}}
		\caption{\label{figEfficiency} \textbf{Measuring efficiency in the loaded device.} (a) System schematic with an axle attached to the central three particles (blue spheres). Red particles are the boundary. (b--d) Three behaviours when the device works against an external load torque, $\tau$. The boundary is driven in the direction of the red arrows. The central three particles rotate in the direction of the blue arrows. (e) Efficiency as a function of torque ratio as measured in simulation for $\pecl=1.75$ (blue) and $\pecl=7$ (red). Dashed lines are the experimentally measured upper bounds on efficiency in the linear response regime. Inset shows efficiency in the low torque limit. Error bars are smaller than the symbols. (f) Experimentally measured gradient of the upper bound on efficiency in the linear response regime as a function of $\pecl$. Line is a guide to the eye. Inset shows the measured angular lag of wall particles behind their optical traps.}
	\end{center}
\end{figure*}

That the slip behaviour depends upon controllable parameters is a striking difference between this softly-coupled rotational system and the rigid rotational elements investigated previously \cite{asavei2013,lin2012,galajda2001,dileonardo2010,xia2010}. Here we can control not only the speed with which the device rotates but also the manner in which this rotation is transmitted through the system. With the development of each successive slip ring, the angular velocity at the centre of the system is reduced relative to that imposed at the boundary and thus the torque transmission is modified and, one would expect, rendered less efficient. In other words, one may say that such soft materials exhibit ``self-lubrication''.

\section{Efficiency}

Having described the rotational behaviour we now consider its efficiency for torque transmission. By controlling the slip behaviour we can tune the transmission of torque from the boundary to the centre. Work is spent through the rotation of the boundary, much of which is dissipated into the surrounding solvent. Nevertheless, since the externally applied torque is transmitted to the innermost particles, useful work can be extracted by attaching an axle, which may be manufactured using commercially available laser lithography nanofabrication techniques \cite{phillips2012optexp,phillips2012epl}. A schematic of this is shown in Fig. 4 (a).

If an external load applies a constant torque, $\tau$, to the axle, the extracted power is $P_\text{o}=\tau \, \omega(\tau)$. Here $\omega(\tau)$ is the angular velocity of the loaded axle. We define the isothermal efficiency as the ratio $\eta=P_\text{o}/P_\text{i}$ of extracted power to input power $P_\text{i}$, \emph{i.e.}, the work spent per unit time \cite{seifert2012,parmeggiani1999}. The latter can be expressed as $P_\text{i}=n \, \avd \, \tod$ with $n=27$ (the number of optically trapped particles), where $\tod$ is the driving torque applied to each of the trapped particles. This definition of efficiency is not unique, but is appropriate here as it relates the mechanical work necessary to move the outer particles to the torque generated at the centre of the device.

This situation is realised in simulation by applying a constant torque to the central particles in the opposite direction to the imposed boundary rotation. The points in Figure 4 (e) show the efficiency obtained from simulations at $N=48$ as a function of the torque ratio for $\pecl=1.75$ (blue) and $\pecl=7$ (red). Efficiency is maximised at an optimal load torque. Further loading reduces the efficiency until the device stalls and the extracted power is zero. Here there is, on average, no rotation of the central particles, while the softness of the assembly allows the outer particles to slip. Beyond the stall torque, the central particles rotate in the opposite direction to the driven particles as the loading dominates the driving. These behaviours are illustrated in Fig. 4 (b--d) and Supplementary Movies 4 (transmission), 5 (stall) and 6 (slip dominated). At $\pecl = 1.75$ (blue data) the system exhibits interior rigid rotation while at $\pecl=7$ (red data) two slip rings exist. This qualitative change in the angular velocity profile manifests as a relative reduction in the efficiency of torque transmission from the boundary to the system centre.

In the experiments, there is no axle and hence $\tau=0$ and all the input power is dissipated. However, from the measured angular velocities in the innermost particle layer, $\omega$, we can determine the efficiency in the linear response regime for small $\tau$, which provides an upper bound $\eta\leq\tfrac{1}{n} (\omega(\tau=0)/\avd)(\tau/\tod)$. Here $\tod$ is calculated from the average angular lag of wall particles behind their optical traps, $\theta$, as $\tod = k R^2 \langle \sin \theta \rangle$. The derivation of these quantities is presented in the Methods. The gradient of this experimentally measured upper bound on $\eta$ is shown as a function of $\pecl$ for $N=48$ in Fig. 4 (f), while the inset shows $\langle \sin \theta \rangle$, which increases linearly with $\pecl$. The dashed lines in Fig. 4 (e) are these upper bounds for the two simulated P\'{e}clet numbers. As demonstrated in the inset, we find excellent agreement with the simulation data at small $\tau$. Although the measured efficiencies are small, it should be noted that efficiencies of a comparable order of magnitude are common in micro- and nanoscale systems in which thermal fluctuations are important. For instance, swimming efficiencies of $\sim 10^{-5}$ are reported for a model self-propelled diffusiophoretic particle \cite{sabass2012}.

\section{Mechanism of transmission control}

\begin{figure*}[htb]
	\begin{center}
		\centerline{\includegraphics[width=130mm]{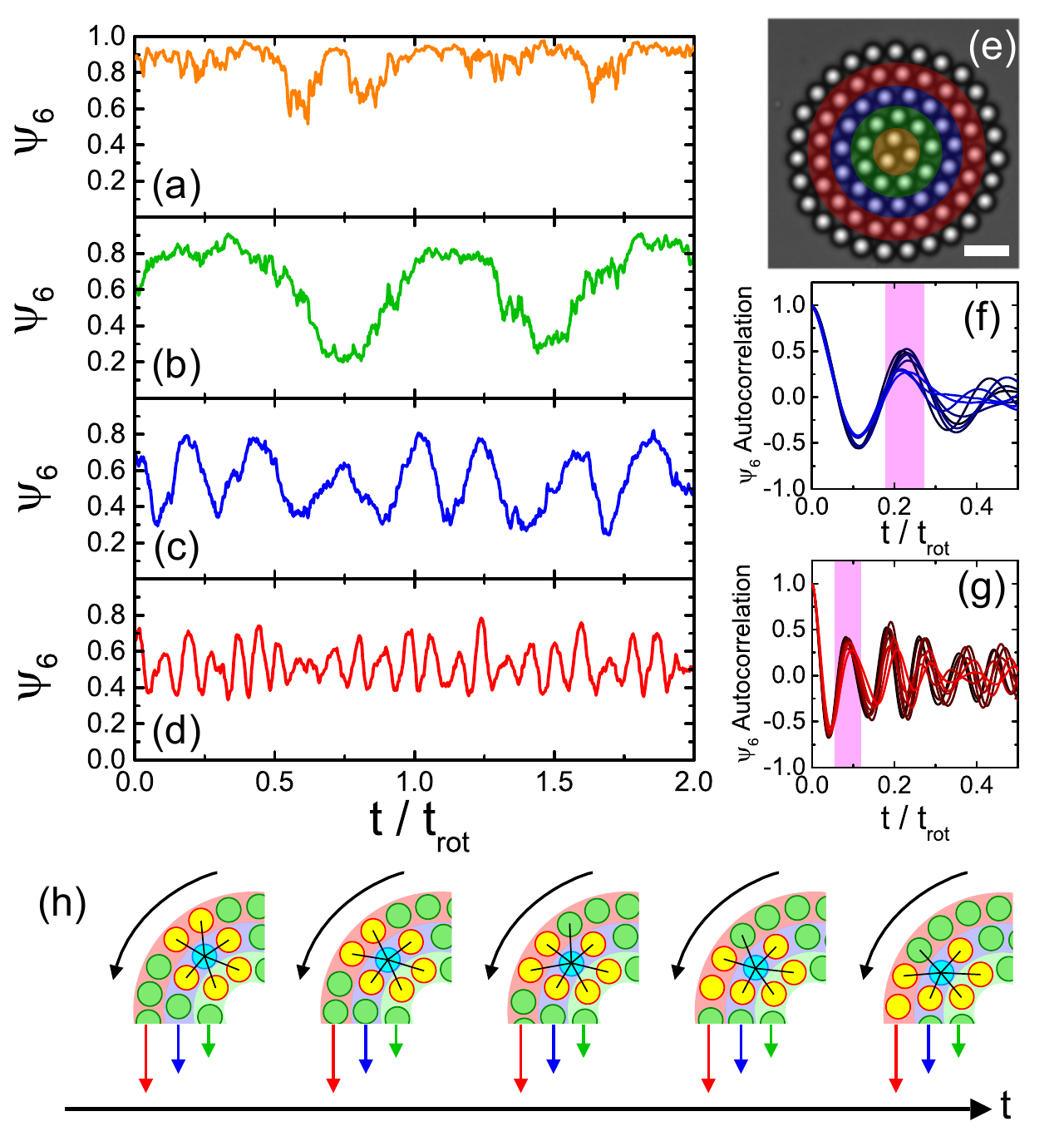}}
		\caption{\label{figPsi6layers} \textbf{Structural fluctuations.} (a -- d) Time dependence of $\psi_6$ in a single experiment at $N=48$ and $\pecl = 19.25$ resolved within layers indicated in (e). (a) Central layer. (b) Layer $2$. (c) Layer $3$. (d) Layer $4$, adjacent to the boundary. (e) Micrograph indicating the identification of particle layers. Scale bar is $10 \; \mathrm{\mu m}$. (f, g) $\psi_6$ autocorrelation for all slipping samples at $N=48$ in layer $3$ (f) and layer 4(g). (h) Illustration of local structural changes due to outer layer overtaking inner layer.
		}
	\end{center}
\end{figure*}

Our particle-resolved experiments allow us to pinpoint the transmission control mechanism. In the rigid and interior-rigid regimes, local structure in the confined assembly is maintained during rotation. However, when layers slip past one another, the local environment around a given particle is in constant flux as its nearest neighbours change. Restricting attention to a single experiment with $N=48$ driven at $\pecl = 19.25$, variations in local structure can be seen in the behaviour of average $\psi_6$ with time within each of four particle layers. These are shown in Fig. 5 (a--d) where the orange, green, blue and red lines correspond to the highlighted layers in Fig. 5 (e). Time is scaled by the rotation period, $t_{\mathrm{rot}}$. In all but the central region, $\psi_6$ explores a large range, indicating that the local structure is alternately driven between highly hexagonal and disordered arrangements ---  rotating the boundary drives the internal structure between the two bistable configurations observed in the quiescent system \cite{williams2013,williams2014}. Furthermore, in all but the central region, this occurs with some characteristic frequency --- fastest in the wall-adjacent layer, and more slowly in the third and second layers.

By considering the time autocorrelation of fluctuations in $\psi_6$ about its mean, the periodicity in $\psi_6$ in each layer is extracted. Figures 5 (f) and (g) show these autocorrelation functions for all slipping data with population $N=48$ in layer 3 (f) and layer 4 (g) --- the blue and red regions shown in (e). When time is scaled by $t_{\mathrm{rot}}$, a common first peak develops in these functions for all samples exhibiting at least two slips indicating a common period for fluctuations in $\psi_6$ at timescale of $\sim 0.08 t_{\mathrm{rot}}$ in layer 4 and $\sim 0.23 t_{\mathrm{rot}}$ in layer 3. No such common peak is observed in layers 2 or 1 and as such the data are not shown. These timescales are interpreted as those of structural self-similarity within these two layers. For the range of rotation speeds explored these timescales are independent of $\pecl$ when time is rescaled by $t_{\mathrm{rot}}$ and are identified as the time taken for a given particle to be overtaken by its faster-moving neighbour in the outer adjacent layer.

Consider the angular velocity of layer 4, $\omega_4 \approx 0.5 \omega_{\mathrm{drive}}$ (see Fig. 2 (b)). In the laboratory frame, the angular velocity of the wall is defined as $\omega_{\mathrm{drive}} = 2 \pi$ radians per $t_{\mathrm{rot}}$. In a reference frame that co-rotates with layer 4, however, the wall moves with angular velocity $\omega_{\mathrm{wall}}^{(4)} = \omega_{\mathrm{drive}} - 0.5 \omega_{\mathrm{drive}} = 0.5 \omega_{\mathrm{drive}} = \pi$ radians per $t_{\mathrm{rot}}$. The wall consists of $27$ particles, and thus the angular displacement required for a wall particle to move one place around the circumference of the circle is $\delta \phi = 2\pi /27 \approx 0.074 \pi$ radians. In the layer 4 co-rotating frame, the time for a wall particle to undergo this angular displacement while moving with constant angular velocity $\omega_{\mathrm{wall}}^{(4)}$ is $\delta \phi / \omega_{\mathrm{wall}}^{(4)} = 0.074 t_{\mathrm{rot}}$. This is remarkably close to the experimentally measured periodicity of $0.08 t_{\mathrm{rot}}$. Performing a similar calculation for the behaviour of layer 3, and noting that layer 4 contains 21 rather than 27 particles predicts a periodicity of $0.024 t_{\mathrm{rot}}$, which is again very similar to the measured timescale of $0.023 t_{\mathrm{rot}}$. The accuracy of these predictions suggests that, in a given layer, this behaviour is dominated by the faster moving outer neighbouring layer. While the layer in question will eventually overtake its slower moving inner neighbouring layer, this process occurs over a much longer timescale and thus the periodicity depends primarily on the angular velocity of the outer layer. Furthermore, we find no change in this periodicity in simulations with torque applied, suggesting that the extraction of work at the centre does not affect the slip behaviour nearer the boundary.

The overtaking process is illustrated in Fig. 5 (h) which shows how the local environment around the cyan particle evolves when particle layers slip over one another. This particle initially has $6$ neighbours, drawn in yellow. As rotation proceeds, the outermost layer overtakes its inner neighbouring layer resulting in a new set of neighbours for the cyan particle. As part of this process the cyan particle briefly has $7$ neighbours, suppressing its $\psi_6$. This is ultimately the origin of the fluctuations observed in Fig. 5 (c) and (d). Given sufficient time, the cyan particle will overtake its inner neighbour, resulting in its $5$-fold co-ordination and again a suppression of $\psi_6$. In other words, the device exhibits a spatial-temporal periodicity related to layers slipping past one another. This provides a slip mechanism which ultimately leads to stalling and breakdown of transmission when the central particles are sufficiently loaded. Interpreted another way, this mechanism may be thought of as relaxation in shearing a glassy system under strong confinement. The fact we can measure the forces on the tweezed particles may offer ways to directly investigate theories of the glass transition~\cite{yoshino2012}.

\section{Discussion}

It is possible to exploit the properties of colloidal assemblies to provide a transmission control mechanism reminiscent of a clutch, which is unique to such soft materials. Although here we employed micron-sized colloidal particles we emphasise that the behaviour we report should remain essentially unaltered provided the constituent particles are large enough that the solvent can be treated as a continuum, \textit{i.e.} around 10 nm in size. Our investigations underline the need to carefully characterise the behaviour of materials in far-from equilibrium conditions if they are to be employed in the fabrication of working engines at small lengthscales. 

We expect this kind of controllable transmission from rigid-body (solid-like) to slipping (fluid-like) behaviour to be generic to a range of soft materials with a suitable yield stress, including protein gels, hydrogels and granular matter. Our work represents an important step in the fabrication of mechanical devices such as gearboxes at the nanoscale and illustrates the application of soft matter physics, developed to tackle colloidal systems, to device design. We expect that devices where torque control is essential~\cite{yim2007} may come to rely upon the controllable transmission mechanisms we present.

\vspace{2mm}
\noindent
\textbf{Acknowledgements}
We acknowledge Jens Eggers, Marco Heinen and Rob Jack for helpful discussion. CPR and IW acknowledge the Royal Society and European Research Council (ERC Consolidator Grant NANOPRS, project number 617266). Additionally, IW was supported by the Engineering and Physical Sciences Research Council (EPSRC). The work of ECO and HL was supported by the ERC Advanced Grant INTERCOCOS (project number 267499). ECO was also supported by the German Research Foundation (DFG) within the Postdoctoral Research Fellowship Program  (project number OG 98/1-1).

\vspace{2mm}
\noindent
\textbf{Author contributions}
I.W. and C.P.R. conceived the experiments. I.W. built the experimental apparatus and performed the experiments. E.C.O., C.P.R and H.L. conceived the simulations. E.C.O. carried out the simulations. T.S. performed the theoretical efficiency analysis. All authors contributed to the data analysis and writing of the manuscript.

\vspace{2mm}
\noindent
\textbf{Additional information}
Correspondence and requests for material should be addressed to H.L. or C.P.R.

\vspace{2mm}
\noindent
\textbf{Competing financial interests}
The authors declare no competing financial interests.

\section*{Methods}

\subsection*{Experimental Details}

We employ a quasi-two-dimensional colloidal sample of polystyrene spheres of diameter $\sigma = 5 \; \mathrm{\mu m}$ and polydispersity $s=2 \%$ suspended in a water--ethanol mixture at a mass ratio of $3:1$. Under these conditions the particles are slightly charged and interact with via a screened electrostatic repulsion with Debye length $\lambda_{\mathrm{D}} \approx 25 \; \mathrm{nm}$. The gravitational length of these particles is $l_g/ \sigma = 0.015 \pm 0.001$, resulting in gravitational confinement to a monolayer adjacent to the glass coverslip substrate in an inverted microscope. This coverslip is treated with Gelest Glassclad 18 in order to prevent particle adhesion.

Twenty-seven particles are manipulated using holographic optical tweezers \cite{bowman2013}. These traps are well-approximated by parabolae with spring constant $k=420(5) \; k_B T \sigma^{-2}$, extracted by observing the constrained Brownian motion of particles in optical traps \cite{williams2013}. Timescales are set by the Brownian diffusion time, empirically determined as $\tau_B \approx 70.2 \; \mathrm{s}$. Particle trajectories are extracted \cite{crocker1996} for $N = 48$ and $N=44$ corresponding to effective area fractions $\phi \approx 0.74$ and $\phi \approx 0.79$.

The boundary is rotated by translating the array of optical traps around a circular path in discrete steps of length $\sigma / 8$. Rotation speed is defined by the frequency of these steps, and is characterised by the P\'{e}clet number, which is the ratio of the Brownian diffusion time to the time taken to drive a boundary particle a distance $\sigma$ along its circular path. We report data for P\'{e}clet numbers in the range $1.75 \le \pecl \le 19.25$ corresponding to optical trap step rates between $0.1$ and $1.1 \; \mathrm{s}^{-1}$. Once boundary rotation is initiated, the system undergoes at least $5$ full rotations before data are acquired. Images are acquired at a rate of $2$ per second for up to $3$ hours.

\subsection*{Simulation Details}
Our Brownian dynamics simulations assume the particles interact via a Yukawa pair potential
\begin{equation}
V(r)=V_0\frac{\mathrm{e}^{-\kappa r}}{\kappa r} ,
\label{eq:yukawa}
\end{equation}
with $r$ denoting the inter-particle separation, $\kappa$ the inverse Debye screening length, and $V_0$ the magnitude of the potential energy. Additionally, each of the 27 particles in the outermost boundary layer are exposed to an harmonic potential mimicking the optical traps employed in experiment, given by
\begin{equation}
V_t (|\textbf{r}_i-\textbf{r}_{i,0}|)= \frac{1}{2} k (|\textbf{r}_i-\textbf{r}_{i,0}|)^2 ,
\label{eq:traps}
\end{equation}
where $\textbf{r}_i$ is the position of $i$th particle and $\textbf{r}_{i,0}$ is the center of its potential well,  with $k$ denoting the trap strength. Each timestep $\delta t$, the locations of the $27$ harmonic potential minima, $\textbf{r}_{i,0}$, are translated a predetermined arc length, $l$, along the boundary resulting in a rotation velocity $l / \delta t$. The velocity, and thus P\'{e}clet number, is controlled by altering this arc length.

In order to account for the hydrodynamic effect of the planar substrate that is present in experiment we apply Blake's solution \cite{vonhansen2011,blake1971} which uses the method of images to obtain the Green's function of the Stokes equation satisfying the no-slip boundary condition at $z=0$ as\begin{equation}
G_{\alpha\beta}(\textbf{r}_i, \textbf{r}_j) = G^S_{\alpha\beta}(\textbf{r}) - G^S_{\alpha\beta}(\textbf{R}) + G^D_{\alpha\beta}(\textbf{R}) - G^{SD}_{\alpha\beta}(\textbf{R}) ,
\label{eq:stokesgreens}
\end{equation}
where the indices $i$ and $j$ refer to spatial positions with $\alpha,\beta=x,y,z$ being the coordinates. The vector $\textbf{r}$ gives the relative position $\textbf{r}_i-\textbf{r}_j$, whereas $\textbf{R}=\textbf{r}_i-{\textbf{r}_j}^{\prime}$ denotes the relative position vector with the $j$'s image at ${\textbf{r}_j}^{\prime}=(x_j,y_j,-z_j)$ with respect to the planar substrate at $z=0$. Blake's solution in Eq. \ref{eq:stokesgreens}, relating the flow field at $\textbf{r}_i$ to a unit point force at $\textbf{r}_j$ in the presence of a boundary plane, comprises four contributions. Firstly, the Green's function for the Stokeslet $G^S$ (\textit{i.e.}, if a unit point force is applied at the origin in an unbounded fluid)
\begin{equation}
G^S_{\alpha\beta}(\textbf{r}) = \dfrac{1}{8\pi\eta}\left(\dfrac{\delta_{\alpha\beta}}{r} + \dfrac{r_{\alpha} r_{\beta}}{r^3}\right) ,
\label{eq:stokeslet}
\end{equation}
with $r=|\textbf{r}|$, and $\eta$ being the viscosity of the fluid. The second contribution is the image Stokeslet. The third is a Stokes doublet $G^D$
\begin{equation}
G^D_{\alpha\beta}(\textbf{r}) = \dfrac{1}{8\pi\eta}2z^2_j(1-2\delta_{\beta z}) \left(\dfrac{\delta_{\alpha\beta}}{r^3} 
			       - \dfrac{3r_{\alpha} r_{\beta}}{r^5}\right) ,
\label{eq:doublet}
\end{equation}
and the final term is a source doublet $G^{SD}$
\begin{equation}
\label{eq:sourcedoublet}
\begin{split}
G^{SD}_{\alpha\beta}(\textbf{r}) = \dfrac{1}{8\pi\eta}2z_j(1-2\delta_{\beta z}) \left(\dfrac{\delta_{\alpha\beta} r_z}{r^3} 
                               - \dfrac{\delta_{\alpha z} r_{\beta}}{r^3} + \right. \\  \left. \dfrac{\delta_{\beta z} r_{\alpha}}{r^3} 
			       - \dfrac{3 r_{\alpha} r_{\beta} r_z}{r^5}  \right) .
\end{split}
\end{equation}

By applying Blake's scheme to the 27 rotating particles located at $\textbf{r}_j$, the $\alpha$th component of the flow field 
reads
\begin{equation}
u_{\alpha=x,y} (\textbf{r}_i) = \sum_{j=1}^{27} \sum_{\beta = x,y} G_{\alpha \beta} (\textbf{r}_i, \textbf{r}_j) f_{\beta}(\textbf{r}_j) ,
\label{eq:flowfield}
\end{equation}
where $\textbf{f}=\gamma \dot{\textbf{r}}_j$ is taken as the drag force applied at $\textbf{r}_j$ caused by the rotation of the boundary, with $\gamma$ being the Stokesian friction coefficient. Note that particle $z$-coordinates are fixed at $2z/\sigma = 1.03$, based on the gravitational length of the experimental colloids. Hence, we restrict our system to effective two-dimensions with vanishing force and flow components along the $z$-direction. The confined particles are further coupled to the flow field via the Stokes' drag 
$\textbf{F}_d (\textbf{r}_i) = \gamma \textbf{u} (\textbf{r}_i)$, with $\textbf{u}$ having components $u_{\alpha=x,y}$ 
as given in Eq. \ref{eq:flowfield}.  

The equation for the trajectory $\textbf{r}_i$ of particle $i$ undergoing Brownian motion in a time step $\delta t$ reads as
\begin{equation}
\textbf{r}_i(t+\delta t) = \textbf{r}_i(t) + \frac{D_0}{k_BT}\textbf{F}_i(t) \delta t + \textbf{u}(\textbf{r}_i) \delta t + \delta \textbf{W}_i  ,
\label{eq:trajectory}
\end{equation}
where $D_0 = k_BT/\gamma$ denotes the free diffusion constant, $k_BT$ the thermal energy, and $\textbf{F}_i(t)$ is the total conservative force acting on particle $i$ stemming from the pair interactions, $V$, and for the 27 driven wall particles additionally from the harmonic trap potential $V_t$. The third term on the right hand side is the solvent flow field due to hydrodynamic effects, the details of which are given in Eq. \ref{eq:flowfield}. The random displacement $\delta \textbf{W}_i$ is sampled from a Gaussian distribution with zero mean and variance $2D_0 \delta t$ (for each Cartesian component) fixed by the fluctuation-dissipation relation.

We use the standard velocity Verlet integration to obtain the equation of motion for the particle trajectories, which contains the total conservative force acting on particles (stemming from the pair interactions and, for the 27 driven particles additionally from the harmonic trap potential), the random displacements (sampled from a Gaussian distribution obeying the fluctuation--dissipation relation) and the determined flow field. Colloid motion is coupled to the flow field via the Stokes' drag.

The simulated unit lengthscale is set by $\kappa$, the energy scale by $k_BT$, and the time scale by $\tau_{\mathrm{sim}} = 1/(\kappa^2 D_0)$. The inverse screening length $\kappa$ is chosen as $\kappa \sigma = 30$, where the experimental value of $\sigma$ serves as a reference. Consequently, the wall radius is $\kappa R_0=130$ yielding the experimental ratio of $R_0 / \sigma = 4.33$. The high screening at $\kappa \sigma = 30$ together with the contact potential chosen as $V(r=\sigma) \approx 1.6 k_BT$ ensures the quasi hard-disc-behaviour. Another crucial parameter in our system is the trap strength which has been set to $k = 0.42 \kappa^2 k_BT$ in order to mimic the measured optical trap stiffness in the experiments. The time step is chosen as $\delta t = 10^{-4} \tau_{\mathrm{sim}}$. Simulations run for up to $3 \times 10^4 \tau_{\mathrm{sim}}$, corresponding to approximately $ 133.3 \tau_B$ with $\tau_B$ being the experimental Brownian time.

\subsection*{Local Structure}

The degree of local hexagonal ordering is quantified using the bond-orientational order parameter, $\psi_6^j = \left| 1/z_j \sum_{m=1}^{z_j} \exp \left(i 6 \theta_m^j \right) \right|$ where $z_j$ is the co-ordination number as defined by a Voronoi construction and $\theta_{m}^{j}$ is the angle made by the bond between particle $j$ and its $m \mathrm{th}$ neighbour and a reference axis. The curved boundary suppresses $\psi_6^j$ in its vicinity. Thus we characterise the hexagonality of a configuration by averaging $\psi_6^j$ over all confined particles that are non-adjacent to the boundary. Where $\psi_6$ is considered within particle layers, $\psi_6^j$ is instead averaged over all particles in the layer in question.

\subsection*{Efficiency calculation}

For completeness, here we provide a derivation of the isothermal efficiency. The total potential energy of the device is
\begin{equation}
  U(\{\x_i\},t) = \sum_{i=1}^n \frac{k}{2}|\x_i-\x_{i,0}(t)|^2 + U_0(\{\x_i\}),
\end{equation}
where the first contribution models the $n=27$ optical traps through harmonic potentials with stiffness $k$ and the second term accounts for the interaction energy of all particles. The potential energy is a function of all particle positions. It becomes explicitly time-dependent since we manipulate the positions of the trap centers
\begin{equation}
  \label{eq:trap}
  \x_{i,0}(\vhi) = R \left(
    \begin{array}{c}
      \cos(2\pi i/n+\vhi) \\ \sin(2\pi i/n+\vhi)
    \end{array}\right)
\end{equation}
through changing $\vhi$ ($R$ is the distance of the traps from the origin). Although this is done stepwise in the experiments and simulations, for simplicity in the following we assume a constant angular velocity $\avd=\dot\vhi$.

By moving the trap centres, the system is driven into a non-equilibrium steady state. The first law of thermodynamics dictates that the work spent to maintain this steady state is balanced by the dissipated heat and the change of internal energy (the signs are convention)~\cite{seifert2012},
\begin{equation}
  \dbar q = \dbar w - \dd U.
\end{equation}
The symbol $\dbar$ stresses the fact that work and heat are not exact differentials. Moreover, they are fluctuating quantities. The work is given by the energy change that is directly caused by the translation of the traps, $\dbar w=\dot w \dd t$, with
\begin{equation}
  \dot w(\{\x_i\}) = \partial_t U = \dot\vhi\sum_{i=1}^n f_i
\end{equation}
and projected forces
$f_i=-k(\x_i-\x_{i,0})\cdot(\partial_\vhi\x_{i,0})$. Plugging in Eq.~\eqref{eq:trap}, the mean work rate becomes
\begin{equation}
  \label{eq:P}
  \mean{\dot w} = n \, \avd \, R^2 \, k \, \mean{\sin\theta} = n \, \avd \, \tod \geqslant 0,
\end{equation}
where $\theta_i=2\pi i/n+\vhi-\vhi_i$ is the lag angle of the $i$th particle behind the trap center. Clearly, its average is independent of the particle index and, moreover, in equilibrium $\mean{\theta}=0$ and therefore $\dot
w=0$. Note that the dynamics (and in particular hydrodynamic interactions) enter only through the average. We extract the lag angles and their average from both the experiments and simulations.

We identify $P_\text{i}=\dot w$ with the input power. Applying a load torque $\tau$, the extracted power reads $P_\text{o}=\tau\omega$, where $\omega(\tau)$ is the (load-dependent) angular velocity of the axle. The
isothermal efficiency finally is defined as the ratio~\cite{parmeggiani1999} 
\begin{equation}
  \eta = \frac{P_\text{o}}{P_\text{i}} =
  \frac{1}{n}\frac{\omega}{\avd}\frac{\tau}{\tod}
\end{equation}
of output to input power.

\newpage
\clearpage
\section*{Supplementary Material}

\setcounter{figure}{0}
\renewcommand{\thefigure}{S\arabic{figure}}

\appendix*
\subsection*{Altering P\'{e}clet Number \textit{in situ}}

\begin{figure}[htb]
	\begin{center}
		\centerline{\includegraphics[width=80mm]{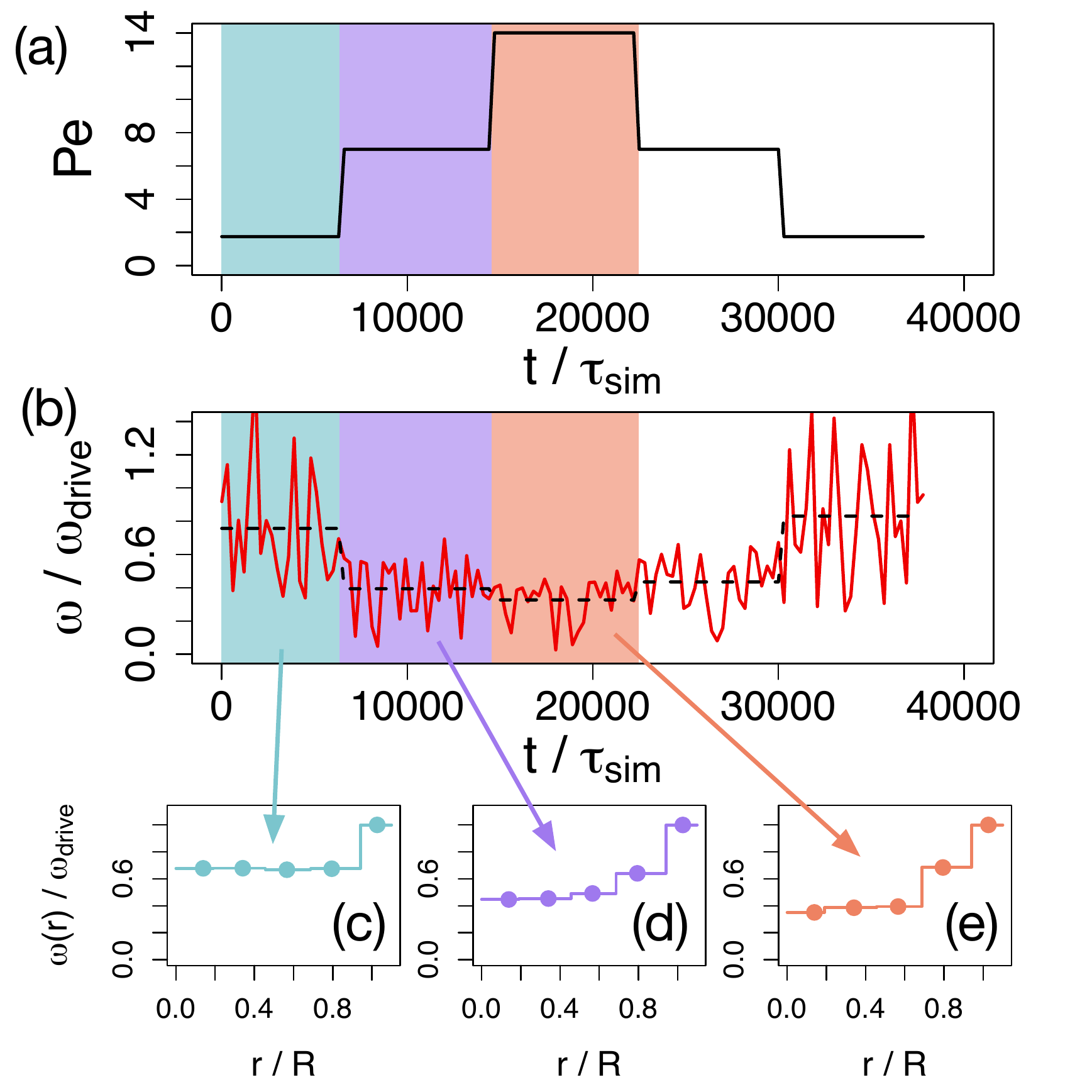}}
		\caption{\label{figSwitchingPe} The effect of altering P\'{e}clet number \textit{in situ} in a single simulation at $N=48$. (a) P\'{e}clet number as a function of time in simulated time units. (b) Relative angular velocity in time interval $300 \tau_{\mathrm{sim}}$ measured in central layer (red line) and average over each region of constant $\pecl$ (black dashed line). (c -- e) Angular velocity profiles in three shaded intervals  corresponding to (c) $\pecl = 1.75$, (d) $\pecl=7$ and (e) $\pecl=14$ }
	\end{center}
\end{figure}

Figure \ref{figSwitchingPe} illustrates that the rotational behaviour of our system can be controlled \textit{in situ} by altering the rate at which the boundary is driven. In a single simulation at $N=48$ the P\'{e}clet number is increased in a stepwise manner from $\pecl=1.75$, through $\pecl=7$ up to $\pecl=14$ and subsequently back down to the $\pecl=1.75$ again. This is shown in Fig. \ref{figSwitchingPe} (a). Panel (b) shows the relative angular velocity measured in the central layer in a time interval of $300 \tau_{\mathrm{sim}}$. The noise in these quasi-instantaneous angular velocity measurements arises due to the small number of particles in the central layer. However, in spite of this it is clear that increasing $\pecl$ results in a relative decrease in the angular velocity of the central layer and thus a change in the transmission mechanism. This is further illustrated in the full angular velocity profiles measured in the three shaded regions shown in (c), (d) and (e), clearly showing a transition from interior-rigid rotation at $\pecl=1.75$ to the development of additional slip between layers as the boundary is rotated more quickly. This represents a further method by which torque transmission may be controlled in soft machines in addition to the alteration of density illustrated in Fig. 3 of the main manuscript.\\

\begin{figure*}[htb]
	\begin{center}
		\centerline{\includegraphics[width=120mm]{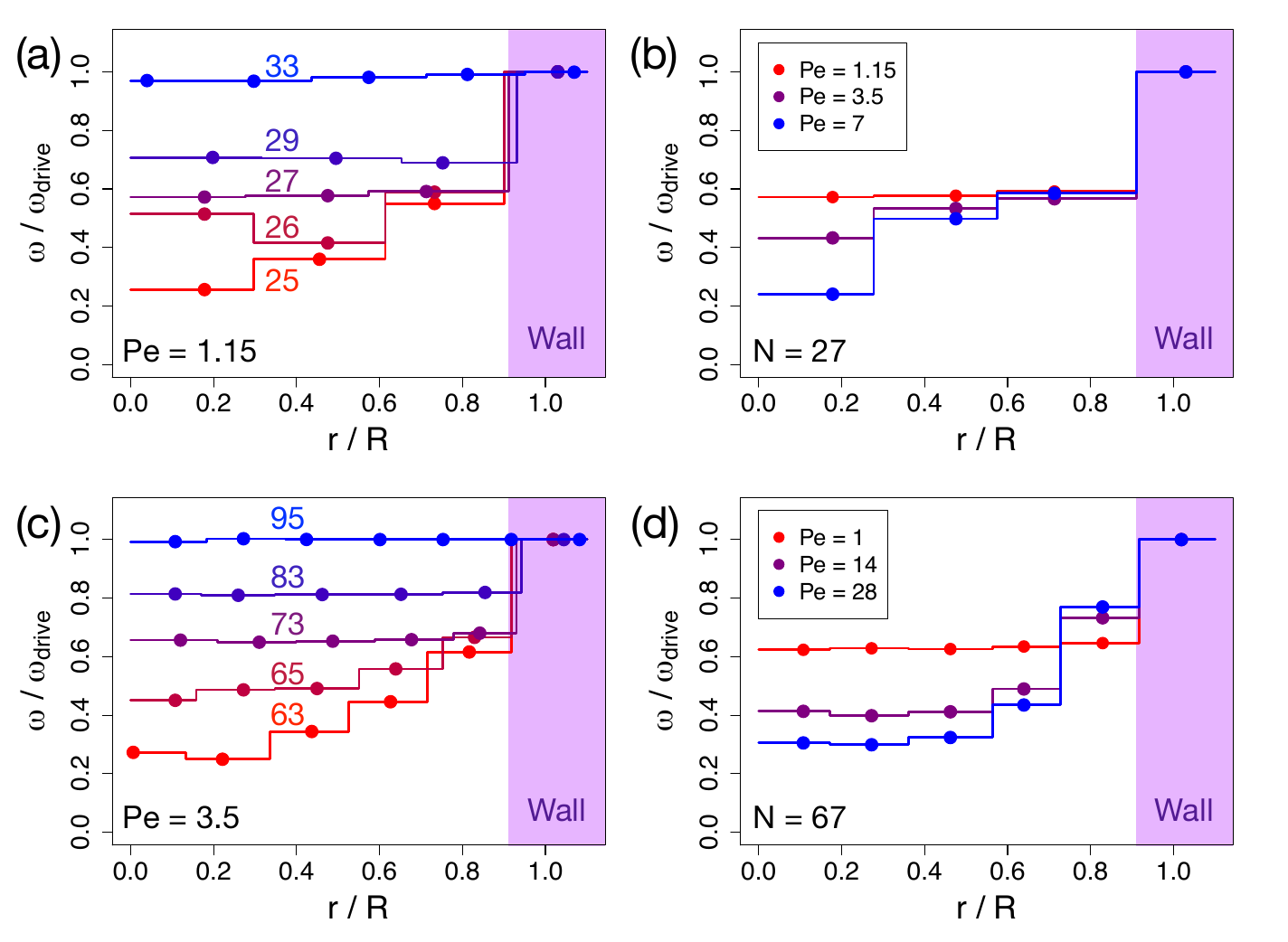}}
		\caption{\label{fig3layer5layer} Angular velocity profiles obtained from simulation of smaller and larger rotating systems. (a) Varying population at constant P\'{e}clet number for a smaller system containing 3 particle layers. Numbers label confined population, $N$. (b) Varying P\'{e}clet number at constant $N = 27$ in the smaller system. (c) Varying population at constant P\'{e}clet number for a larger system containing 5 particle layers. Numbers label confined population, $N$. (d) Varying P\'{e}clet number at constant $N=67$ in the larger system.}
	\end{center}
\end{figure*}

\subsection*{Rotational behaviour of larger and smaller systems}

In order to show that the rotational behaviour reported for our experimental and simulated system is general to both larger and smaller systems additional simulation are performed for with $n=21$ and $n=33$ driven boundary particles resulting in systems containing $3$ and $5$ confined particle layers. Supplementary Fig. \ref{fig3layer5layer} shows angular velocity profiles for the smaller (a, b) and larger (c, d) systems analogous to those shown in Fig. 2 of the main article. The change in the angular velocity profiles when population is varied at constant $\pecl$ (a, c) is qualitatively identical in both the larger and smaller system, with rigid body rotation observed at high population and slip developing between particle layers as density is decreased. Similarly, at constant population, both the smaller and larger system exhibit the development of slips as $\pecl$ is increased (b, d).

\end{document}